\documentclass{article}
\usepackage{spconf,amsmath,graphicx}
\usepackage{booktabs}
\usepackage{multirow}
\usepackage{amssymb}
\usepackage{pifont}
\usepackage{hyperref}
\title{MLCA-AVSR: Multi-Layer Cross Attention Fusion based Audio-Visual Speech Recognition}
\name{He Wang$^1$, Pengcheng Guo$^1$, Pan Zhou$^2$, Lei Xie$^{1{\ast}}$ \thanks{*Corresponding author}}
\address{$^1$Audio, Speech and Language Processing Group (ASLP@NPU), School of Computer Science,\\Northwestern Polytechnical University, Xian, China\\$^2$Space AI, Li Auto}
\begin{document}
\ninept
\maketitle
\begin{abstract}
While automatic speech recognition (ASR) systems degrade significantly in noisy environments, audio-visual speech recognition (AVSR) systems aim to complement the audio stream with noise-invariant visual cues and improve the system's robustness. 
However, current studies mainly focus on fusing the well-learned modality features, like the output of modality-specific encoders, without considering the contextual relationship during the modality feature learning. 
In this study, we propose a multi-layer cross-attention fusion based AVSR (MLCA-AVSR) approach that promotes representation learning of each modality by fusing them at different levels of audio/visual encoders. 
Experimental results on the MISP2022-AVSR Challenge dataset show the efficacy of our proposed system, achieving a concatenated minimum permutation character error rate (cpCER) of 30.57\% on the Eval set and yielding up to 3.17\% relative improvement compared with our previous system which ranked the second place in the challenge. 
% After multi-system fusion, our proposed system outperforms the first-place system, with a new SOTA cpCER of 29.13\% on this dataset.
Following the fusion of multiple systems, our proposed approach surpasses the first-place system, establishing a new SOTA cpCER of 29.13\% on this dataset.
\end{abstract}
\section{Introduction}
\label{sec:intro}
With the rapid advancement of artificial intelligence, automatic speech recognition (ASR) systems have made considerable progress in recognition accuracy and even reached human parity~\cite{xiong2016achieving}. 
However, in complex acoustic environments or real-world far-filed scenarios like multi-person meetings, ASR systems can be susceptible to performance degradation due to background noise, inevitable reverberation, and multiple speaker overlap. 
Conversely, visual-based speech recognition (VSR) systems are immune to acoustic environments or distorted speech signals~\cite{hong2023watch,ma2023auto}. 
As a result, there is growing interest among researchers in incorporating visual features into ASR models to mitigate the impact of damaged speech signals, leading to the emergence of audio-visual speech recognition (AVSR).

A considerable amount of research on AVSR has shown that integrating visual information into ASR models can enhance the robustness in complex acoustic environments significantly. 
In~\cite{ma2021end}, Ma \textit{et al.} proposed an end-to-end dual-encoder hybrid CTC/Attention~\cite{kim2017joint} AVSR system, which includes a ResNet~\cite{he2016deep} based visual encoder, a Conformer~\cite{gulati2020conformer} based audio encoder, and a multi-layer perception (MLP) module to fuse different modality features. 
In contrast to the MLP-based fusion strategy, Sterpu \textit{et al.}~\cite{sterpu2018attention} first introduced an attention-based fusion mechanism and found that aligning the features of different modalities can improve the learned representations. 
Following this, plenty of studies have adopted a cross-attention module to capture inherent alignments and complementary information between fully encoded audio-visual representations~\cite{wei2020attentive,wu2021audio,zhang2023ve}. 
Additionally, some works directly concatenate the raw speech and video sequences together and employ a shared encoder with self-attention mechanisms to learn modality alignments~\cite{hong2023watch,li2023robust}. 
In~\cite{cheng2022dku,li2023xmu}, hidden features from different layers of audio and visual encoders were leveraged to achieve more effective fusion, indicating that conducting multi-layer fusion can promote the performance of AVSR systems.

Recently, the Multi-modal Information based Speech Processing (MISP) Challenge series~\cite{chen2022audio,chen2022first,wang2023multimodal} has been introduced to explore the utilization of both audio and visual data in distant multi-microphone signal processing tasks, like keyword spotting and speech recognition. 
In the audio-visual diarization and recognition (AVDR) track of the MISP2022 Challenge, multi-channel audio data and lip reading video data were suggested to use in developing robust speech recognition systems that retain high accuracy in far-field home TV scenarios. 
Among the submitted systems, Xu \textit{et al.}~\cite{xu2023nio} achieved the highest ranking by summing audio and visual features and performing channel-wise attention between the fusion features and multi-channel audio signals. 
Guo \textit{et al.}~\cite{guo2023npu} proposed a single-layer cross-attention (SLCA) fusion based AVSR system, which employed a cross-attention module to combine the features of different modalities. 
However, both approaches conducted modality fusion based on fully encoded audio and visual representations, without considering the fusion during the representation learning phase. 
Although Li \textit{et al.}~\cite{li2023xmu} explored the usage of multi-level modal features, the simple concatenation based fusion did not effectively capture the alignments between different modalities.

In this paper, we propose a multi-layer cross attention fusion based audio-visual speech recognition (MLCA-AVSR) model to enhance the robustness of the AVSR model. 
Specifically, our model improves the previous SLCA fusion module and integrates it into multiple intermediate layers of the modality encoders. 
By fusing at different intermediate layers, our model allows each modality to learn complementary contextual information from the other modality, ranging from low-level to high-level features and from fine-grained details to abstract global patterns. 
This MLCA module enables a more exhaustive representation learning and fusion process. 
Furthermore, we also employ the Inter-CTC~\cite{lee2021intermediate} loss to guide the output feature of each cross-attention module. 
To the best of our knowledge, this is the first attempt to integrate the cross-attention module into the intermediate layer of modality encoders and simultaneously conduct modality fusion during representation learning. 
% We conduct comprehensive experiments on the MISP2022-AVSR challenge dataset~\cite{chen2022audio}, evaluating the performance of different audio and visual encoders and validating the effectiveness of the proposed audio-video fusion module.
% The results demonstrate that our MLCA-AVSR system surpasses the first-place system in the MISP2022 challenge, achieving a new state-of-the-art result on this dataset.
The experiment results on the MISP2022-AVSR challenge dataset~\cite{chen2022audio} demonstrate that our MLCA-AVSR system surpasses the first-place system in the MISP2022 challenge, achieving a new SOTA result on this dataset.

\begin{figure}[tb]
  \centering
  \centerline{\includegraphics[height=7cm]{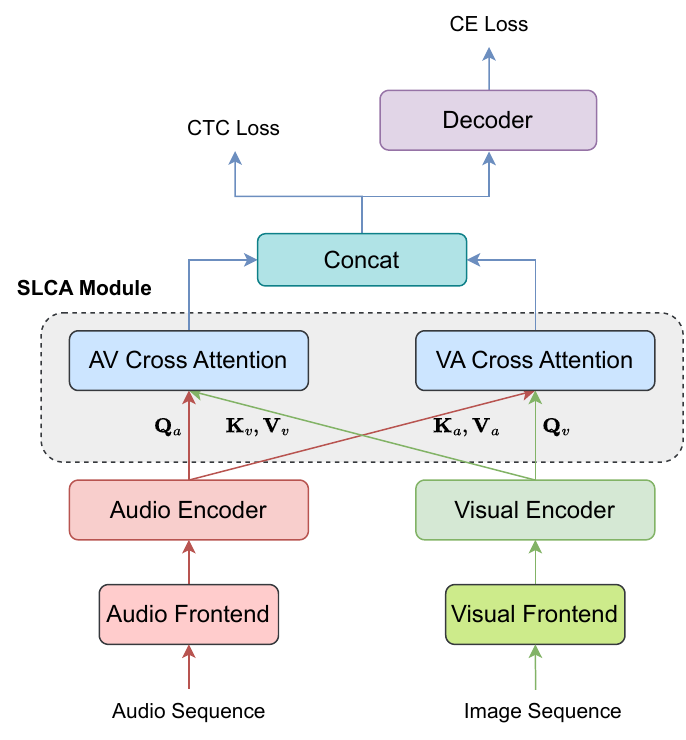}}
\caption{An overview of our previous system in~\cite{guo2023npu}.}
\label{fig:1}
\end{figure}

\section{Method}
\subsection{Previous System}
\label{sec:previous-system}
Fig.\ref{fig:1} shows the overall framework of our previous system~\cite{guo2023npu}, which mainly consists of four components: audio/visual frontends, audio/visual encoders, an SLCA fusion module, and a decoder. 
In detail, both audio and visual data are first processed through frontend modules for feature extraction. 
The visual frontend consists of a 5-layer ResNet3D module implemented with 3D convolutions, while the audio frontend utilizes a structure composed of 2 convolutional down-sampling layers. 
After the frontend modules, the obtained features are separately passed through two modality-specific Branchformer~\cite{peng2022branchformer} encoders, which model the input features into well-learned representations. 
The SLCA fusion module includes two modality-based cross-attention layers. 
In each layer, one modality's representation is used as the Query vector, while the other modality's representation serves as the Key and Value vectors. 
Take the cross-attention of the audio branch (``AV Corss-Attention" in Fig.~\ref{fig:1}) as an example, the computation can be formulated as:
\begin{equation}
\operatorname{Attention}(\mathbf{Q}_a, \mathbf{K}_v, \mathbf{V}_v)=\operatorname{Softmax}\left(\frac{\mathbf{Q}_a\ \mathbf{K}_{v}^\mathbf{T}}{\sqrt{d_k}}\right)\ \mathbf{V}_v, 
\end{equation}
where the subscript $a$ refers to audio representations and $v$ for visual representations, respectively. 
The $d_k$ represents the attention dimension. 
With the help of the cross-attention fusion layers, each modality can learn relevant and complementary contextual representation from the other modality. 
Finally, the features of different modalities are concatenated together to calculate the connectionist temporal classification (CTC)~\cite{graves2006connectionist} loss and fed into a Transformer~\cite{vaswani2017attention} decoder to generate predicted tokens and compute the cross-entropy (CE) loss. 
For our AVSR system, we advance the SLCA module introduced in~\cite{guo2023npu} and integrate it into the intermediate layers of modality encoders to promote representation learning for each modality. 
In the following section, we will begin with the improved cross-attention module and then provide a detailed description of the MLCA-AVSR system.

\begin{figure}[htb]
  \centering
  \centerline{\includegraphics[height=7cm]{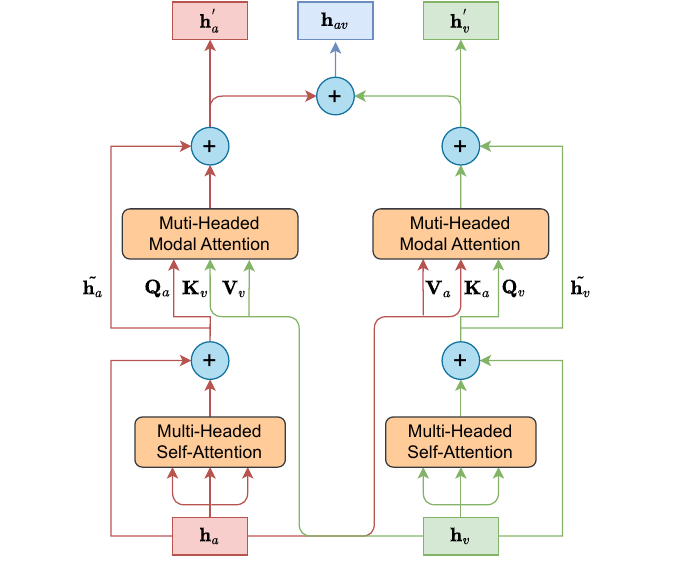}}
\caption{An overview of the improved cross-attention module.}
\label{fig:2}
\end{figure}

\subsection{Improved Cross Attention}
As illustrated in Fig.~\ref{fig:2}, our improved cross-attention module comprises two main parts, which are the audio feature processing flow and the visual feature processing flow. 
Both parts are composed of a multi-headed self-attention layer~\cite{vaswani2017attention} and a multi-headed modal-attention layer, with a residual connection applied around each of them. 
Given the audio features $\mathbf{h}_a$ and visual features $\mathbf{h}_v$, both of them are first fed into the multi-headed self-attention layer to obtain deeper representations and summed with their respective original input through a residual connection, resulting in intermediate representations $\tilde{\mathbf{h}_a}$ and $\tilde{\mathbf{h}_v}$. 
Subsequently, two multi-headed modal-attention layers are employed: the audio multi-headed modal-attention (AMMA) in the audio feature processing flow and the visual multi-headed modal-attention (VMMA) in the visual feature processing flow. 
In the AMMA, the Query $\mathbf{Q}_a$ is generated from the audio intermediate representation $\tilde{\mathbf{h}_a}$, while the Key $\mathbf{K}_v$ and Value $\mathbf{V}_v$ are derived from the video input $\mathbf{h}_v$. 
Conversely, in the VMMA, the Query $\mathbf{Q}_v$ is generated from the intermediate video representation $\tilde{\mathbf{h}_v}$, while the Key $\mathbf{K}_a$ and Value $\mathbf{V}_a$ are transformed from the audio input $\mathbf{h}_a$. 
The outputs of the AMMA and VMMA are then combined with their respective intermediate representations through residual connections, resulting in the audio flow output $\mathbf{h}_{a}^{'}$ and video flow output $\mathbf{h}_{v}^{'}$. 
Finally, the output of both flow are added together to obtain the fused audio-visual feature of the cross-attention module, denoted as $\mathbf{h}_{av}$. 
In summary, the computation of the cross-attention module can be formulated as:
\begin{equation}
    \tilde{\mathbf{h}_a}= \mathbf{h}_a + \operatorname{MHSA}(\mathbf{h}_a),
\end{equation}
\vspace{-0.65cm}
\begin{equation}
    \tilde{\mathbf{h}_v}= \mathbf{h}_v + \operatorname{MHSA}(\mathbf{h}_v),
\end{equation}
\vspace{-0.5cm}
\begin{equation}
\tilde{\mathbf{h}_a}\xrightarrow{\operatorname{linear}}\mathbf{Q}_a;\mathbf{h}_v\xrightarrow{\operatorname{linear}}\mathbf{K}_v,\mathbf{V}_v,
\end{equation}
\vspace{-0.5cm}
\begin{equation}
\tilde{\mathbf{h}_v}\xrightarrow{\operatorname{linear}}\mathbf{Q}_v;\mathbf{h}_a\xrightarrow{\operatorname{linear}}\mathbf{K}_a,\mathbf{V}_a,
\end{equation}
\vspace{-0.5cm}
\begin{equation}
    \mathbf{h}_{a}^{'}= \tilde{\mathbf{h}_a} + \operatorname{AMMA}(\mathbf{Q}_a,\mathbf{K}_v,\mathbf{V}_v),
\end{equation}
\vspace{-0.5cm}
\begin{equation}
    \mathbf{h}_{v}^{'}= \tilde{\mathbf{h}_v} + \operatorname{VMMA}(\mathbf{Q}_v,\mathbf{K}_a,\mathbf{V}_a),
\end{equation}
\vspace{-0.5cm}
\begin{equation}
    \mathbf{h}_{av}= \mathbf{h}_{a}^{'} + \mathbf{h}_{v}^{'},
\end{equation}
where $\operatorname{MHSA}$ means multi-headed self-attention module and $\xrightarrow{\operatorname{linear}}$ represents linear projection. 
\label{sec:proposed-system}

\subsection{Multi-Layer Cross Attention Fusion}

\begin{figure}[htb]
  \centering
  \centerline{\includegraphics[width=0.9\linewidth]{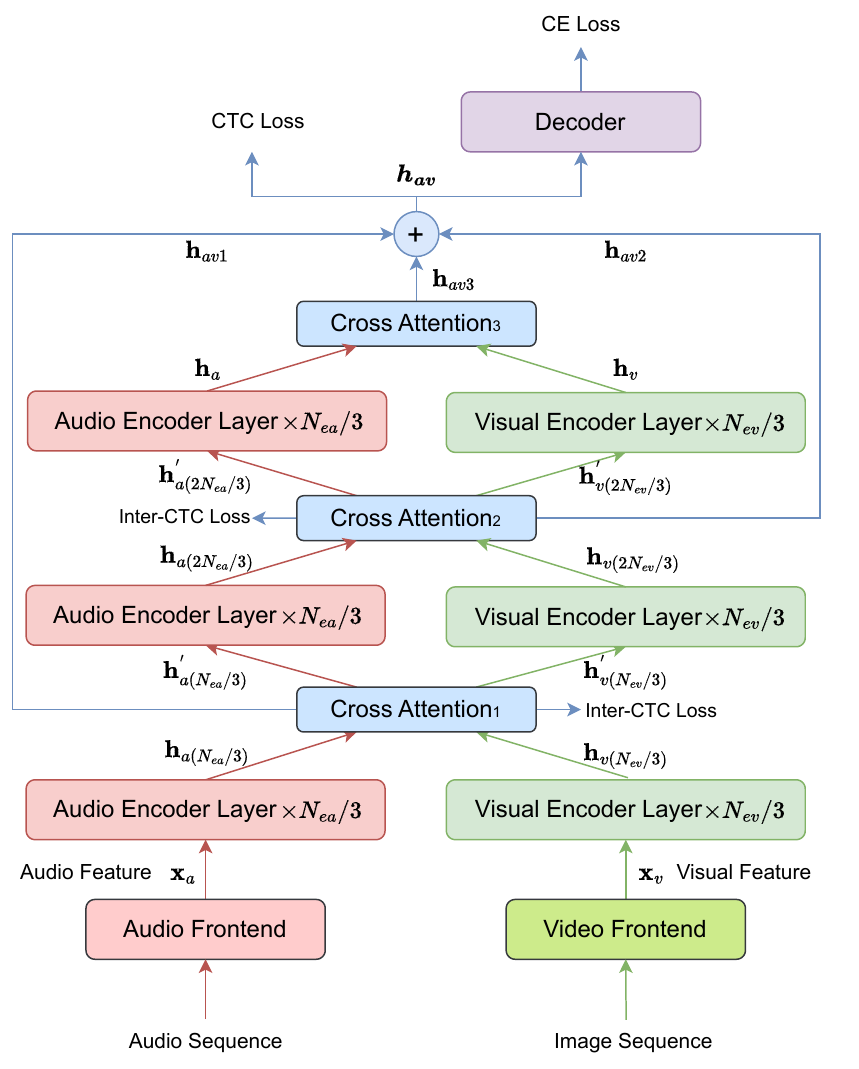}}
\caption{An overview of proposed MLCA-AVSR model}
\label{fig:3}
\end{figure}

Fig \ref{fig:3} depicts the structure of our MLCA-AVSR model. It consists of four main components: audio and visual frontends, audio and visual encoders, a fusion module, and a decoder. 
Similar to our previous system described in Section~\ref{sec:previous-system}, we adopt a 2-layer convolutional down-sampling network as the audio frontend and a ResNet3D network as the video frontend. 
For the audio and visual encoder, we adopt the recently proposed E-Branchformer architecture~\cite{kim2023branchformer}, which has demonstrated superior performance compared with the Branchformer. 
The decoder component remains the Transformer and the model is optimized with the joint CTC/Attention~\cite{kim2017joint} training strategy.

The MLCA-based fusion module is significantly improved over the SLCA system. 
It introduces two additional cross-attention modules within the audio and visual encoders, distributed evenly across multiple layers. 
This allows for a better fusion of different modalities by effectively leveraging the hidden representations in the encoders. 
We define the number of audio and visual encoder layers as $N_{ea}$ and $N_{ev}$, respectively. 
The inputs of the audio and visual encoder are denoted as $\mathbf{x}_a$ and $\mathbf{x}_v$. 
$\mathbf{h}_{a(i)}$ and $\mathbf{h}_{v(i)}$ represent the output of the $i$-th layer of the audio and visual encoder, respectively. 
$\mathbf{h}_{a(iN_{ea}/3)}^{'}$, $\mathbf{h}_{v(iN_{ev}/3)}^{'}$, and $\mathbf{h}_{avi}$ means the audio feature, video feature, and audio-visual fusion feature outputted by the $i$-th cross-attention module. 
The output audio and video features from the cross-attention modules, located within encoders, are then individually used as inputs to the next layer of the audio and visual encoder. 
The audio-visual fusion features, obtained from each cross-attention module, are added to form the final output $\mathbf{h}_{av}$ of the MLCA module.

The outputs of $\text{cross-attention}_1$ and $\text{cross-attention}_2$ are used as intermediate outputs of MLCA-AVSR to calculate the Inter-CTC losses, which guide the cross-attention modules to better utilize the hidden representations of encoders for audio-visual feature fusion.

\section{Experiment}
\label{sec:experiment}
\subsection{Data Processing}
\textbf{Dataset.} 
All model training and testing experiments are conducted on the MISP2022-AVSR dataset~\cite{chen2022audio}, a large-scale audio-visual Chinese conversational corpus consisting of 141h audio and video data collected by far/middle/near microphones and far/middle cameras in 34 real-home TV rooms. 
% Specifically, during the data recording, the 2-channel middle microphone array and middle cameras are 1.5 to 3 meters away from the speakers, while the 6-channel far microphone array and far camera are 3 to 5 meters away.s
% In addition, each speaker has a near earphone and a middle camera right in front of them. 
After segmented by the oracle diarization, it contains 106.09 hours of dialogue audio and video in the training set, 3.09 hours in the development set, and 3.13 hours in the evaluation set. \\
\textbf{Audio Processing.} 
% Fig.~\ref{fig:1} demonstrates the procedure of the data processing. 
Initially, both the Middle and Far data are pre-processed by the WPE~\cite{yoshioka2012generalization} and GSS~\cite{boeddeker2018front} algorithms to effectively extract enhanced clean signals for each speaker. 
Subsequently, the combination of the enhanced data and original Near data is speed perturbed with 0.9, 1.0, and 1.1 factors. 
To simulate realistic acoustic environments, we employ the MUSAN~\cite{snyder2015musan} corpus and the open-source pyroomacoustics toolkit\footnote{https://github.com/LCAV/pyroomacoustics} to generate authentic background noises and room impulse responses (RIRs). 
The combined dataset for training comprises approximately 1300 hours, encompassing both the processed and simulated data. \\
\textbf{Video Processing.} 
For the video data, we crop the ROIs corresponding to the lip of the speaker and resize it to $112\times112$. 
During the training process, random rotation, horizontal flipping, and color-jittering strategies are applied to augment the video data. \\
\textbf{Speaker Diarization.} 
% According to the rule of the MISP2022 challenge, the ground-truth diarization is not available for the final Eval set.
In order to compare the performance of our system with the participants' of MISP2022 in the final Eval set without ground-truth diarization, we also use the speaker diarization (SD) model~\cite{pang2022tsup} with a diarization error rate (DER) of 9.43\% on the Dev set, same as our previous system, to segment the Dev and Eval set, denoted as $\text{Dev}^{sd}$ and $\text{Eval}^{sd}$ respectively.
% Specifically, the SD model is implemented with a Transformer diarization module and a speaker embedding module, achieving a diarization error rate (DER) of 9.43\% on the Dev set.

\subsection{Setup}
All of the models, including ASR, VSR, and AVSR, are implemented with the ESPnet toolkit~\cite{watanabe2018espnet}. 
For the audio-based ASR model, we use a 24-layer E-Branchformer as the encoder, each with 256-dim, 4 attention heads, and 1024-dim feed-forward inner linear projection. 
Additionally, the decoder contains 6 Transformer layers, each with 4 attention heads and 2048-dim feed-forward. 
Both the encoder and decoder's dropout rates are set to 0.2. 
For the video-based VSR model, the visual frontend is a 5-layer ResNet3D module, whose channels are 32, 64, 64, 128, 256, and kernel size is 3, the visual encoder is a 9-layer E-Branchformer, and others are the same as ASR systems. 
To validate the superior performance of the E-Branchformer encoder, we also trained Conformer and Branchformer ASR and VSR models with similar model sizes to the E-Branchformer. 
For the audio-visual speech recognition model, the cross-attention module uses 4 attention heads with 256 attention dims. 
At the beginning of training, the audio and visual encoders are initialized using well-trained ASR and VSR models, respectively.

% \begin{figure}[tb]
%   \centering
%   \centerline{\includegraphics[width=8.5cm]{figures/data_processing.pdf}}
% \caption{The flow chart of data processing and simulation. N$\times$ refers to N times the original 106 hours Near data.}
% \label{fig:1}
% \end{figure}

\subsection{Results and Analysis}
\subsubsection{Single-modality ASR and VSR models}
As shown in Table~\ref{table-1}, we conduct performance comparison among Conformer, Branchformer, and E-Branchformer encoders for both ASR and VSR models. 
For Conformer and Branchformer encoders, we expand the model size of ASR and VSR systems by increasing the number of encoder layers or feature dimensions, ensuring the fairness of comparison. 
The results on the MISP2022-AVSR dataset indicate that the E-Branchformer encoder exhibits a performance advantage over Conformer and Branchformer encoders. 
Specifically, E-Branchformer ASR and VSR models achieve 0.8\% and 2.2\% relative CER improvement compared with Branchformer models, respectively, and 1.3\% and 3.8\% relative improvement over Conformer on the Eval set. 
Although the VSR models are immune to noisy environments, they still have fairly high CERs compared to ASR models.

\subsubsection{Comparison to common fusion methods}
To better demonstrate the advantages of the multi-layer cross-attention fusion method, we implement two common fusion methods for comparison: simple addition of the outputs from the audio and visual encoders, and fusion through an MLP after concatenating the outputs along feature dimension. 
The MLP module consists of a two-layer linear projection with 2048-dim and 256-dim, respectively. 
For a fair comparison, we increase the number of E-Branchformer encoder layers for audio and visual encoders in Add and MLP fusion experiments from 24 to 27 and 9 to 12, respectively. 
Table~\ref{table-2} presents the experimental results of audio-visual speech recognition systems utilizing different fusion strategies. 
Clearly, our proposed multi-layer cross-attention (MLCA) fusion method outperforms the Add and MLP, yielding up to 2.4\% and 2.5\% relative CER improvement on the Eval set, respectively.
\begin{table}[tp]
	\centering
	\caption{The CER(\%) results of ASR and VSR systems on Dev and Eval sets segmented by ground-truth speaker diarization, and cpCER(\%) results on Eval set segmented by SD model ($\text{Eval}^{sd}$).}
    \resizebox{\linewidth}{!}{
        \begin{tabular}{cccccc}
    		\toprule
    		Encoder & Modal & Param & Dev & Eval & $\text{Eval}^{sd}$ \\
    		\hline
    		Conformer & \multirow{3}*{A} & 77.64M & 27.3 & 28.6 & 35.4 \\
    		Branchformer & ~ & 79.25M & 26.8 & 28.1 & 33.9 \\
    		E-Branchformer & ~ & 70.78M & \textbf{25.9} & \textbf{27.3} & \textbf{33.3} \\
            \hline
            Conformer & \multirow{3}*{V} & 48.07M & 87.4 & 88.4 & 86.4 \\
            Branchformer & ~ & 41.67M & 85.7 & 86.8 & 84.9 \\
    		E-Branchformer & ~ & 37.59M & \textbf{84.4} & \textbf{84.6} & \textbf{83.5} \\
    		\bottomrule
    	\end{tabular}
    }
	\label{table-1}
\end{table}
\begin{table}[tp]
    \centering
    \caption{The CER(\%) results of Add, MLP, and MLCA fusion AVSR systems on Dev and Eval sets, while cpCER(\%) results on $\text{Eval}^{sd}$ set .}
    \begin{tabular}{ccccc}
        \toprule
        Fusion & Param & Dev & Eval & $\text{Eval}^{sd}$ \\
        \hline
        Add & 110.05M & 24.6 & 26.5 & 32.5 \\
        MLP & 111.63M & 24.9 & 26.6 & 32.8 \\
        MLCA (ours) & 105.40M & \textbf{21.8} & \textbf{24.1} & \textbf{30.6} \\
        \bottomrule
    \end{tabular}
    \label{table-2}
\end{table}

\subsubsection{Ablation experiments}
To further explore the role of each layer in the MLCA module, and whether the cross-attention module utilizing hidden layer features within the audio/visual encoders improves the performance of the AVSR system, we conducted a series of ablation experiments on the three cross-attention modules, namely $\text{cross-attention}_1$, $\text{cross-attention}_2$, and $\text{cross-attention}_3$, as shown in Fig.~\ref{fig:3}. 
The results of ablation experiments are presented in Table \ref{table-3}.
It is evident that removing either the $\text{cross-attention}_1$ or $\text{cross-attention}_2$ module within the audio-visual encoders has a certain impact on the performance of the MLCA-AVSR system. 
Furthermore, retaining one layer of the cross-attention module in the audio-visual encoders also can be beneficial for improving the performance, particularly when retaining the shallower module $\text{cross-attention}_1$ to make audio-visual fusion happen earlier, achieving a relative 0.6\% CER improvement on the Eval set than only retaining $\text{cross-attention}_2$. 
In conclusion, incorporating two cross-attention modules within the audio-visual encoders makes good use of multi-level modal features to fuse audio-visual features, yielding up to 2.3\% CER relative improvement over the system without $\text{cross-attention}_1$ and $\text{cross-attention}_2$ on the Eval set.
\begin{table}[tp]
	\centering
	\caption{The CER(\%) results of cross-attention ablation systems on Dev and Eval sets, while cpCER(\%) results on $\text{Eval}^{sd}$ set. $\text{CA}_1$, $\text{CA}_2$, $\text{CA}_3$ means $\text{cross-attention}_1$, $\text{cross-attention}_2$ and $\text{cross-attention}_3$, respectively.}
        \begin{tabular}{ccccccc}
    		\toprule
            \multicolumn{4}{c}{\textbf{Proposed Method}} & \multirow{2}*{Dev} & \multirow{2}*{Eval} & \multirow{2}*{$\text{Eval}^{sd}$}  \\
            \cmidrule(lr){1-4}
    		$\text{CA}_1$ & $\text{CA}_2$ & $\text{CA}_3$ & Param & ~ & ~ & ~ \\
    		\hline
    		 $ \text{\ding{55}} $ & $ \text{\ding{55}} $ & $\text{\ding{51}}$ & 99.09M & 24.5 & 26.4 & 32.5 \\
    		 $ \text{\ding{55}} $ & $\text{\ding{51}}$ & $ \text{\ding{51}} $ & 102.24M & 23.1 & 25.5 & 31.4 \\
    		 $\text{\ding{51}}$ & $ \text{\ding{55}} $ & $\text{\ding{51}}$ & 102.24M & 22.7 & 24.9 & 31.1 \\
              $\text{\ding{51}}$ & $\text{\ding{51}}$ & $\text{\ding{51}}$ & 105.40M & \textbf{21.8} & \textbf{24.1} & \textbf{30.6} \\
    		\bottomrule
    	\end{tabular}
	\label{table-3}
\end{table}
\begin{table}[tp]
	\centering
	\caption{The cpCER(\%) results of MISP2022 top 3 systems and ours on $\text{Dev}^{sd}$ and $\text{Eval}^{sd}$ set .}
        \begin{tabular}{cccc}
    		\toprule
    		Model & MISP Rank & $\text{Dev}^{sd}$ & $\text{Eval}^{sd}$ \\
    		\hline
    		Channel AV-Fusion~\cite{xu2023nio} & $1^{st}$ & -  & 29.58 \\
            \hline
    		SLCA-AVSR (previous)~\cite{guo2023npu} & \multirow{2}*{$2^{nd}$} & 29.73 & 33.74 \\
            +ROVER & ~ & 28.13 & 31.21 \\
            \hline
    		Conformer-AV~\cite{li2023xmu} & \multirow{2}*{$3^{rd}$} & 29.40 & 33.94 \\
            +ROVER & ~ & 27.50 & 31.88 \\
            \hline
    		MLCA-AVSR (\textbf{ours}) & \multirow{2}*{-} & 25.92 & 30.57 \\
            +ROVER & ~ & \textbf{24.49} & \textbf{29.13} \\
    		\bottomrule
    	\end{tabular}
	\label{table-4}
\end{table}
\subsubsection{Comparison with MISP2022 top systems}
Table \ref{table-4} shows comparison results among our proposed MLCA-AVSR system and the top 3 systems in the AVDR track of the MISP2022 challenge. 
It can be observed that MLCA-AVSR outperforms our previous SLCA-AVSR system, with a relative improvement of 3.81\% and 3.17\% cpCER on the $\text{Dev}^{sd}$ set and $\text{Eval}^{sd}$ set respectively, even surpassing the results of multi-system fusion using ROVER~\cite{fiscus1997post} technique in the SLCA-AVSR, by 0.64\% CER improvement. 
After applying ROVER to the proposed MLCA-AVSR system, the cpCER of the $\text{Eval}^{sd}$ set is reduced by 0.45\% compared to the system ranked first in the MISP2022 challenge, achieving a new SOTA performance of the MISP2022-AVSR dataset.

\section{Conclusion}
\label{sec:conclusion}
This paper proposes a multi-layer cross-attention fusion based AVSR (MLCA-AVSR) system, building upon our previous single-layer cross-attention fusion based AVSR (SLCA-AVSR) system which ranked second in the MISP2022 challenge. 
MLCA-based fusion incorporates two cross-attention modules within encoders to fuse audio-visual representations at different levels. 
On the MISP2022-AVSR dataset, our proposed MLCA-AVSR system outperforms our previous system and surpasses the system ranked first in the challenge after multi-system fusion, achieving a new SOTA result.

% \section{REFERENCES}
% \label{sec:refs}
\bibliographystyle{IEEEbib}
\bibliography{refs}

\end{document}